# Lattice Gas Automata Simulation of 2D site-percolation diffusion: Configuration dependence of the theoretically expected crossover of diffusion regime

Mehrdad Ghaemi[1,2,*], Nasrollah Rezaei-Ghaleh[2,3], Yazdan Asgari[2]

[1] Chemistry Dept., Tarbiat moallem Univ., Tehran, Iran
ghaemi@tmu.ac.ir
[2] Center for Complex Systems Research, K.N. Toosi Univ. Technology, Tehran, Iran
yazdan1130@gmail.com
[3] Inst. Biochem. Biophys., Univ. Tehran, Tehran, Iran
nrezaeig@ibb.ut.ac.ir

**Abstract.** Theoretical analysis of random walk on percolation lattices has predicted that, at the occupied site concentrations of above the threshold value, a characteristic crossover between an initial sub-diffusion to a final classical diffusion behavior should occur. In this study, we have employed the lattice gas automata model to simulate random walk over a square 2D site-percolation lattice. Quite good result was obtained for the critical exponent of diffusion coefficient. The random walker was found to obey the anomalous sub-diffusion regime, with the exponent decreasing when the occupied site concentration decreases. The expected crossover between diffusion regimes was observed in a configuration-dependent manner, but the averaging over the ensemble of lattice configurations removed any manifestation of such crossovers. This may have been originated from the removal of short-scale inhomogeneities in percolation lattices occurring after ensemble averaging.

## 1 Introduction

To treat the static and dynamic properties of systems with inherent disorders, theory of percolation has proven useful in a large variety of areas. Biological evolution, protein diffusion in biological membranes, disease epidemics, forest fires and social phenomena are some relatively new examples of the wide applicability of this theory [1-5]. In spite of this, there exist some purely theoretical challenges in the area and much effort has been dedicated to solve them, with theoretical and computational tools [6].
The static and dynamic properties of site percolation lattices have been extensively investigated during the recent decades using theoretical, computational and even experimental methods [7-9]. It is well known that, as the concentration $P$ of the occupied sites approaches a threshold value of $P_c$, an infinite cluster of the occupied sites over which the unbounded diffusion or conduction can take place is formed [10]. For

*Corresponding author

$P>P_c$, the probability of an occupied site to be on the infinite cluster, $P_\infty$ is given by the characteristic exponent $\beta$ through the scaling formula

$$P_\infty \sim (P-P_c)^\beta \quad \textbf{(1)}$$

while $P_\infty$ is zero for $P<P_c$ [11]. There is also a percolation correlation length, $\xi$, which for length scales $r \geq \xi$, the percolating lattice appears homogeneous, but for $r<\xi$, it exhibits a self-similar fractal geometry. $\xi$ is zero for $P=1$, but as $P$ approaches $P_c$ from the right side, it diverges as [12]

$$\xi \sim (P-P_c)^{-\upsilon} \quad \textbf{(2)}$$

For site-percolation lattices with Euclidean dimension of $d=2$, $P_c$, $\beta$ and $\upsilon$ are known to be 0.592746 [13], 5/36 [11] and 4/3 [12], respectively.
Extensive efforts have been made to study the dynamic properties of percolation lattices such as diffusion or conduction over them and relate these properties to the static characteristics of such lattices. It is self-evident that below $P_c$, diffusion and conduction will be ultimately restricted by the perimeter of finite clusters, but at $P_c<P<1$ where $\xi \neq 0$, it has been conjectured [14,15] that the diffusing (or conducting) particle would undergo an anomalous sub-diffusive behavior, obeying the law

$$<R^2>^{1/2} \propto N^k \quad \textbf{(3)}$$

where $<R^2>^{1/2}$ is the ensemble average of the Euclidean displacement, $N$ is the number of time steps, and $k$ is an exponent equal to 1/2 for classical diffusion and less than 1/2 for the sub-diffusion regime. It is believed that for $<R^2>^{1/2} >> \xi$, the diffusing particle will behave classically (i.e. $k=1/2$), but the diffusion coefficient $D$ will scale through the following relation [16]

$$D \sim (P-P_c)^\mu \quad \textbf{(4)}$$

For $d=2$, the initial value of $k$ occurring when $<R^2>^{1/2} << \xi$ has been numerically calculated as 0.348 [17]. The Alexander-Orbach (AO) conjecture, pointing out that the spectral dimension (defined as $2kD_f$ where $D_f$ is the fractal dimension) is independent of $d$ and equal to 4/3 [18], suggests that the dynamic exponent $\mu$ must be related to the static ones through the following relation:

$$\mu=[(3d-4)\upsilon-\beta]/2 \quad \textbf{(5)}$$

so that for $d=2$, the dynamic exponent $\mu$ should be as 91/72 (i.e. about 1.264) [7,8]. However, the AO conjecture, although remains as a remarkably accurate estimate, is not precisely correct and the true value of spectral dimension is slightly smaller than 4/3 for $d<6$ [8]. Therefore, the dynamic exponent $\mu$ may be different from that proposed by the AO conjecture. In past, a variety of techniques have been employed to estimate $\mu$, including numerical methods, analytical approximations such as series expansions, small cell real-space renormalization technique, ε-expansion method and even experimental techniques and values ranging from 1.20-1.32 have been reported for $\mu$ in $d=2$ [8,16, 19-22]. It is also assumed that the change of diffusion regime occurs after some characteristic crossover time, $\tau$, which diverges at $P=P_c$, so that at the percolation threshold, the diffusion is anomalous for all $N$ [15,16].

This study aimed at simulating random walk on percolation lattices using the lattice gas automata (LGA) approach. LGA are discrete dynamical systems in regard with space, time, and the states of the system. Each point in a regular spatial lattice, called a cell, can have a finite number of particles. The particles in the lattice move according to a local rule. That is, the movement of a particle at a given time depends only on its own state one time step previously, and the states of its nearby neighbors at the previous time step. All cells on the lattice are updated synchronously. Thus the state of the entire lattice advances in discrete time steps. Many LGA are two-dimensional due to visualization and computational concerns, but higher-dimensional lattices certainly exist [23,24]. Using LGA approach, we explored first how the diffusion coefficient ($D$) depends on the concentration of the occupied sites ($P$) in the lattice and obtained the related exponent of $\mu$. Since the obtained value of $\mu$ was reasonably consistent with the literature [8,16,19-22], we were convinced that our model has grasped the essential features of the studied phenomenon. Thereafter, the nature of diffusion regime and its dependence on $P$ was investigated. More specifically, we examined if and how the theoretically-expected transition between anomalous sub-diffusion and classical diffusion regimes occurred. It was found that the cross-over between two regimes occurs in a configuration-dependent manner and the averaging over the ensemble of configurations removes any manifestation of such crossovers. The time required for crossover to occur was also found to be configuration-dependent.

## 2 Results and Discussion

All simulations were conducted with the Lattice Gas Automata (LGA) of the square 2D lattices of size L=1000. This size is longer than the estimated correlation length of the lattice if $P$ is sufficiently higher than $P_c$ ($P$=0.61 and higher), so the lattice seems to be large enough to represent the theoretically-predicted crossover. Lattices of similar size have been previously employed to manifest such crossovers [e.g. 16]. The conducting sites where the random walker was permitted to enter were introduced randomly in specified concentrations, according to a uniform random distribution. Random numbers were generated by Mersenne Twister (MT19937) algorithm [25]. The walker is initially located at the center of the lattice. At each step, the walker must select one of the occupied sites within its von-Neumann neighborhood with equal probabilities to enter. No waiting time is allowed. The transition rule remains unchanged during the whole process of 150,000 time steps.

For each concentration of conducting sites ($P$), the random walk was simulated over 10000 random configurations and the ensemble average of the squared Euclidean displacement of the random walker ($<R^2>$, where $<>$ indicates its ensemble-average) was plotted against time steps ($N$) and the effective diffusion coefficient ($D_{\mathrm{eff}}$) was estimated through calculating the slope of this plot in double linear scale. Fig. 1 displays how $D_{\mathrm{eff}}$ varies as $P$ is changed between 0 and 1. As expected, $D_{\mathrm{eff}}$ was effectively zero at $P$ below a critical value of ($P'_c$) about 0.60 and started to smoothly increase above it. The more accurate examination of the concentration dependence of $D_{\mathrm{eff}}$ between $P$s of 0.50 and 0.60 revealed that $D_{eff}$ was 0.0000 at $P$<0.57, 0.0005 at $P$=0.57, 0.0019 at $P$=0.58, 0.0096 at $P$=0.59 and 0.0193 at $P$=0.60 (see the Inset of Fig.1). It is well known that the percolation threshold $P_c$ is around 0.592746 where the single infinite cluster is first formed [13]. At $P<P_c$, it is expected that the diffusion will ultimately be restricted by the perimeter of finite clusters, so that providing the simulation is long enough, the diffusion coefficient will be effectively zero. The non-zero value of $D_{eff}$ at $P$=0.57-0.59 may be caused by the insufficiency of 150,000 time steps for the random walker to reach at the boundary of the large finite clusters. Thereafter, considering the dynamic scaling law $D \sim (P-P'_c)^{\mu}$, log($D_{\text{eff}}$) was plotted against log($P$-$P'_c$) (Fig. 2) to estimate the value of $\mu$. The calculated $\mu$ was 1.251±0.010, reasonably consistent with the values proposed by Alexander-Orbach conjecture [18] and reported in some references [8,16,19-22]. The obtained value for $\mu$ was found to be nearly independent of lattice size, within the range of $L$=1000-5000.

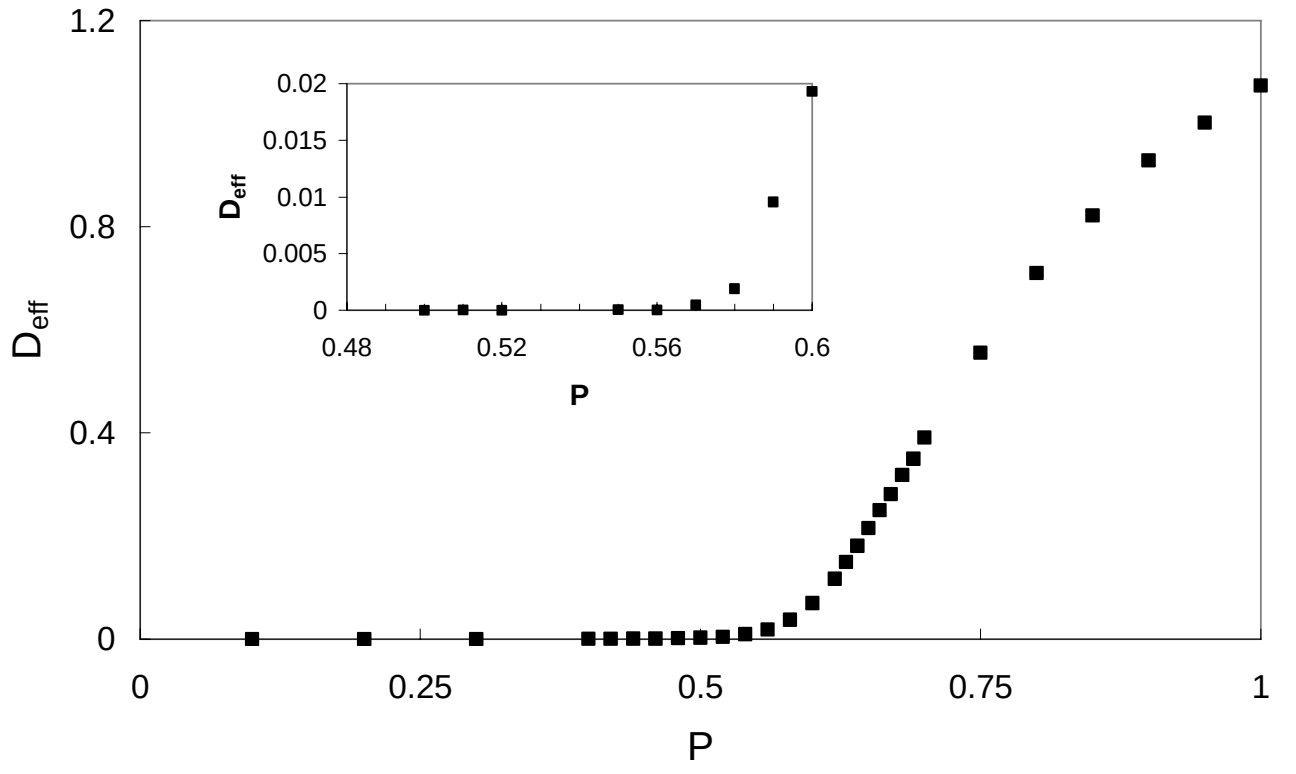

**Fig. 1.** Effective diffusion coefficient ($D_{eff}$) dependence on $P$, the concentration of the occupied sites. Inset shows this dependence for $P$ between 0.50 and 0.60.

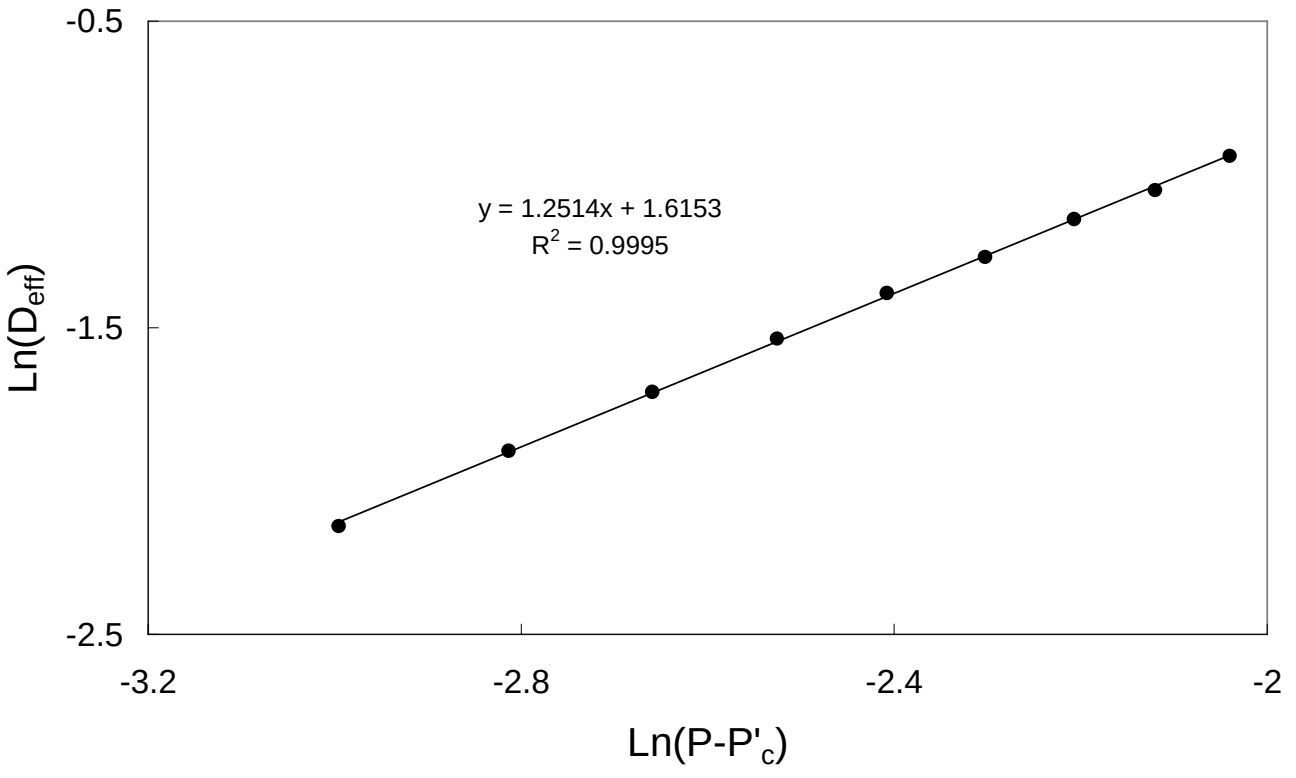

**Fig. 2.** The scaling behavior of $D$ with ($P$- $P'_c$)

To investigate if the random walk on the percolation lattice obeyed the classical or anomalous regime of diffusion, the squared Euclidean displacement of the random walker (<$R^2$>) was plotted against time steps ($N$) in double logarithmic scale for each concentration of conducting sites ($P$) above $P'_c$. As manifested in Fig. 3, log <$R^2$> seems to vary linearly with log *(N)* at all studied $P$s above $P'_c$, although the slopes of the linear plots ($2k$) smoothly decrease with $P$ (see the inset of Fig. 3). Similar to $\mu$, the obtained values for $k$ were insensitive to lattice size within the range of $L$=1000-5000. These results may be taken to indicate that, when $P$ approaches $P'_c$ from the right side, random walk on percolation lattice progressively demonstrates the sub-diffusive behavior. This was in contrast with a fixed value of about 1/3 expected according to the theory [15]. However, at $P$ sufficiently higher than threshold, deviation from the theoretically-expected $k$ of 1/3 and significant dependence on $P$ have been previously reported [e.g. in 16].

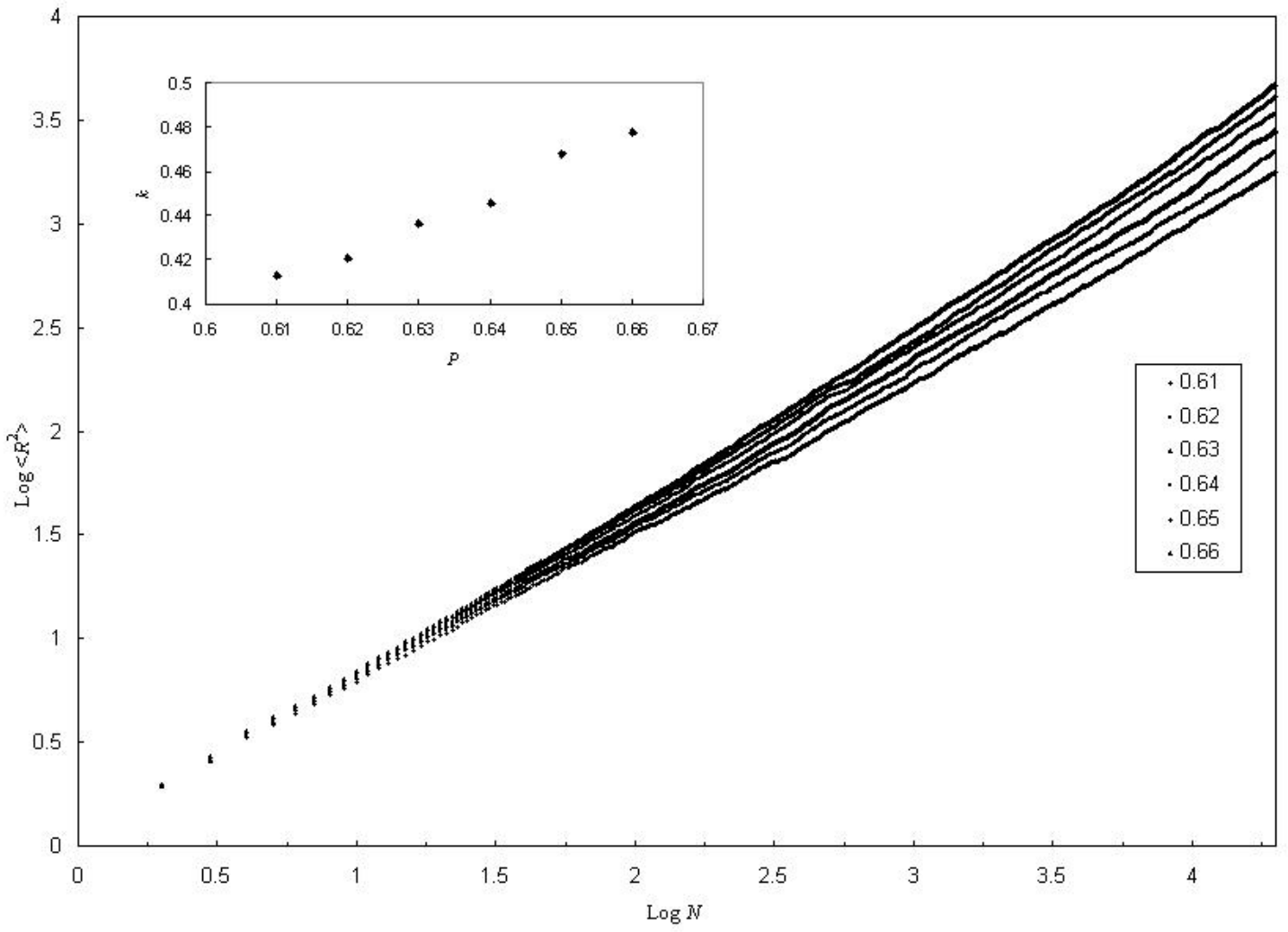

**Fig. 3.** Log-Log plot of the mean square displacement, <$R^2$>, versus $N$ for various concentrations of the occupied sites, $P$. $P$ varies between 0.66 and 0.61 from top to down. The inset shows how $k$ varies with $P$.

Theoretical analysis of anomalous diffusion on percolation lattices have frequently shown that, at $P_c<P<1$ where $\xi\neq 0$ ($\xi$ denotes the correlation length of the lattice) and the fractal cluster structure influences random walk on the lattice [26], the random walker would initially exhibit the anomalous sub-diffusion behavior (with $k<1/2$), but after a characteristic crossover time $\tau$, the random walker would behave classically (with $k=1/2$) [15,16]. To examine if such a crossover could be simulated by our LGA model, the local value of exponent $k$ was estimated for different $N$ from the slope of log $R^2$ versus log *(N)* plot. Figure 4 demonstrates the evolution of $k$ in a wide range of $N$, where $1<\log(N)<5$. In contrast with theoretical prediction, $k$ was found to fluctuate around a nearly constant value along the course of walk at all *P*s examined, and neither the characteristic crossover nor any smooth change from anomalous sub-diffusion to normal diffusion regime was observed.

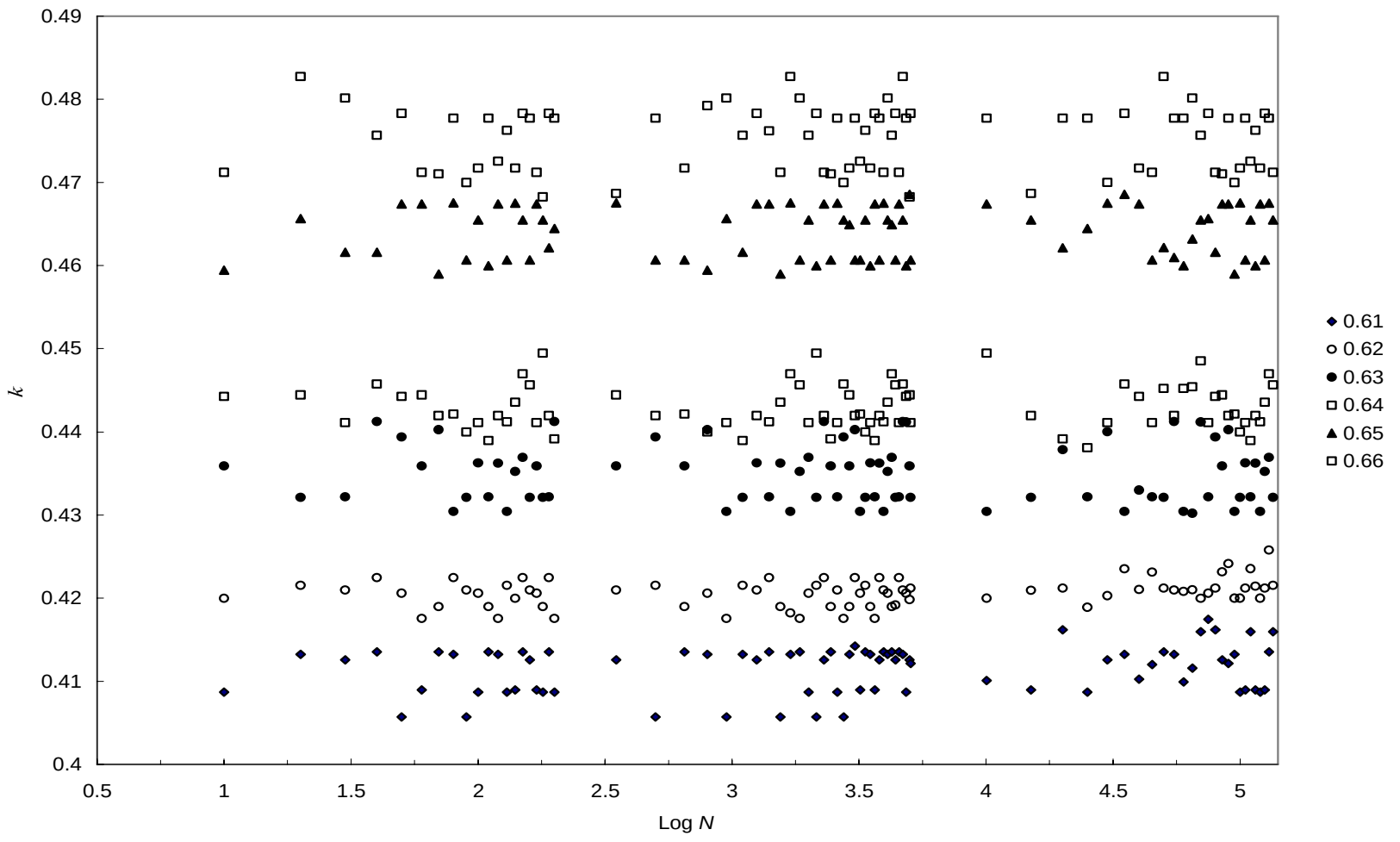

**Fig. 4.** The behavior of *k* with Log*N* for various concentrations of the occupied sites, *P*. *P* varies between 0.66 and 0.61 from top to down.

While no characteristic crossover could be represented in diffusion behavior when the Euclidean displacements were averaged over the ensemble of LGA configurations, it was observed that, at numerous LGA configurations but not all of them, a sharp transition can be found between two regimes of diffusion. Fig. 5A depicts two of the observed crossovers between various diffusion regimes. However, this phenomenon is strongly configuration-dependent and in several LGA configurations, this transition is completely disappeared (see Fig. 5B). Interestingly, the crossover time $\tau$ also revealed prominent configuration dependency. In order to obtain the crossover time $\tau$, we used the interception point of two fitted lines before and after the transition observed in the graph of $R^2$ versus time. As Fig. 6A-B demonstrates, for random configurations generated with specific seeds showing a characteristic crossover, the crossover time $\tau$ steps upward when *P* approaches $P'_c$ from the right side although the extent and accurate position of this upward stepping are variable for different random seed numbers.
Finally, our results may be taken to indicate that the transition between two diffusion regimes is manifested in a configuration-dependent manner and the ensemble averaging over the lattice configurations disappears such transitions. This effect of ensemble averaging has previously been reported for the anisotropic nature of random walk over two-dimensional percolation clusters [27]. This effect may be originated from the removal of short-scale inhomogeneities in percolation lattices due to ensemble averaging.

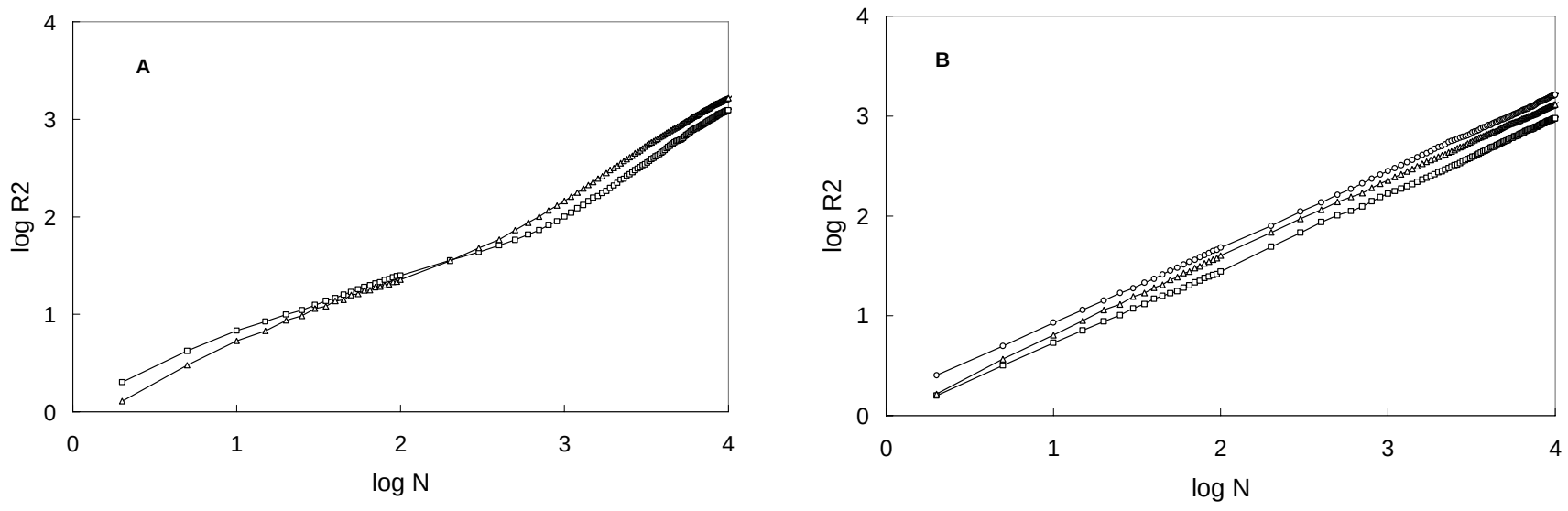

**Fig. 5.** Log-Log plot of $R^2$ versus $N$. A) for lattice configurations generated with random seeds of 1 or 3. B) for lattice configurations generated with random seeds of 2, 19 or 23.

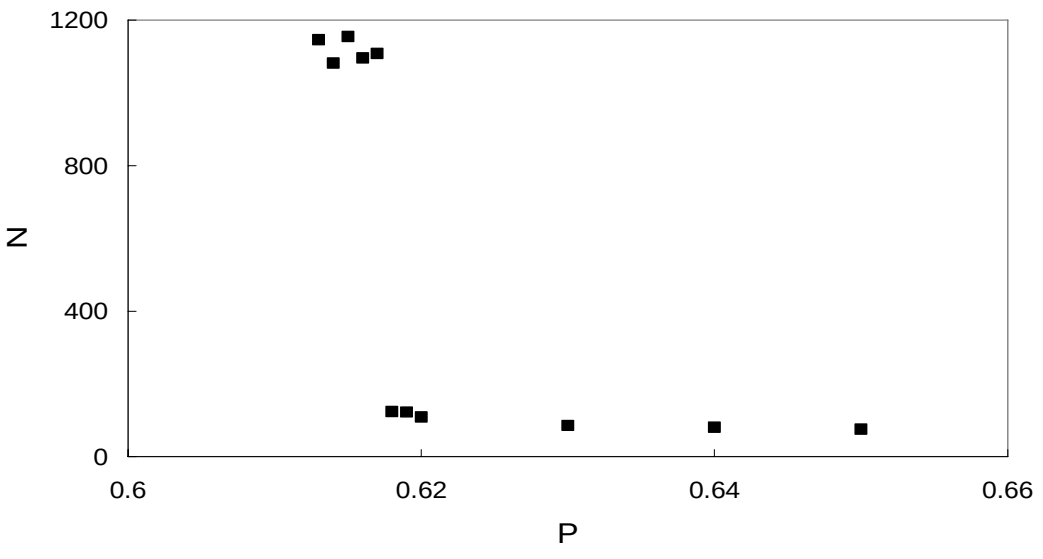

A

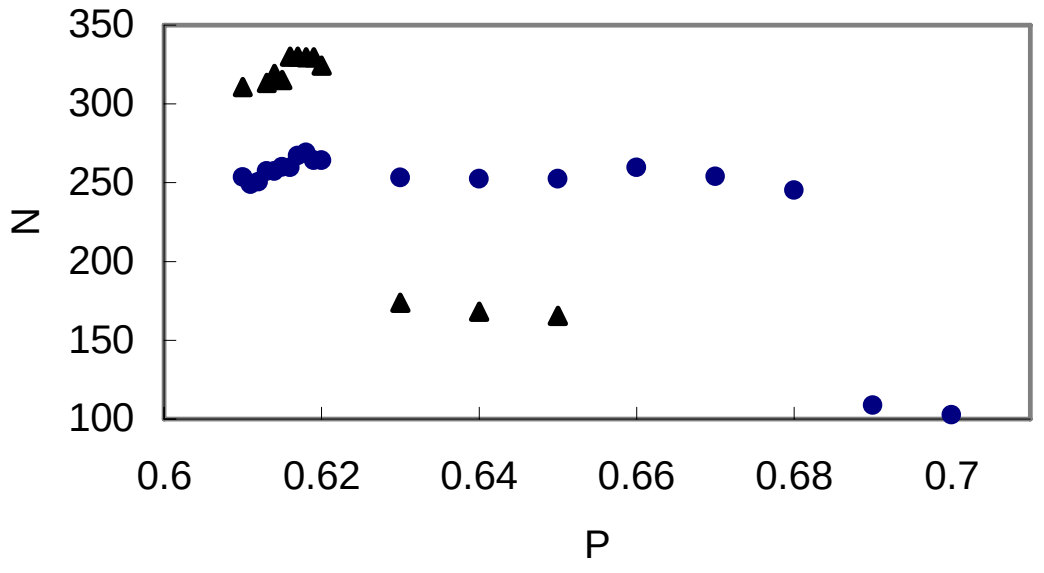

B

**Fig. 6.** The time step $N$ required for crossover to occur decreases sharply when $P$ increases A) For a random configuration generated with a specific seed B) For two other random configurations generated with two distinct seeds

## 3 Conclusion

We have employed the lattice gas automata model to simulate random walk over a square 2D site-percolation lattice. For the critical exponent of diffusion coefficient quite good result was obtained and the random walker was found to obey the anomalous sub-diffusion regime, with the exponent decreasing when the occupied site concentration decreases. The expected crossover between diffusion regimes was observed in a configuration-dependent manner, but the averaging over the ensemble of lattice configurations removed any manifestation of such crossovers. Finally, our results may be taken to indicate that the transition between two diffusion regimes is manifested in a configuration-dependent manner and the ensemble averaging over the lattice configurations disappears such transitions which may have been originated from the ensemble averaging.